# ESTIMATING HIGH-DIMENSIONAL INTERVENTION EFFECTS FROM OBSERVATIONAL DATA

BY MARLOES H. MAATHUIS, MARKUS KALISCH AND PETER BÜHLMANN

*ETH Zürich*

We assume that we have observational data generated from an unknown underlying directed acyclic graph (DAG) model. A DAG is typically not identifiable from observational data, but it is possible to consistently estimate the equivalence class of a DAG. Moreover, for any given DAG, causal effects can be estimated using intervention calculus. In this paper, we combine these two parts. For each DAG in the estimated equivalence class, we use intervention calculus to estimate the causal effects of the covariates on the response. This yields a collection of estimated causal effects for each covariate. We show that the distinct values in this set can be consistently estimated by an algorithm that uses only local information of the graph. This local approach is computationally fast and feasible in high-dimensional problems. We propose to use summary measures of the set of possible causal effects to determine variable importance. In particular, we use the minimum absolute value of this set, since that is a lower bound on the size of the causal effect. We demonstrate the merits of our methods in a simulation study and on a data set about riboflavin production.

**1. Introduction.** Our work is motivated by the following problem in biology. We want to know which genes play a role in a certain phenotype, say a disease status or, in our case, a continuous value of riboflavin (vitamin $B_2$) production in the bacterium *Bacillus subtilis*. To be more precise, our goal is to infer which genes have an effect on the phenotype in terms of an intervention. If we knocked down single genes, which of them would show a relevant or important effect on the phenotype? The difficulty is, however, that the available data are only observational. For our concrete problem, we observe the logarithm of the riboflavin production rate as a continuous response and expression measurements from essentially the whole genome of *B. subtilis* as high-dimensional covariates. Using such observational data, we want to infer all (single gene) intervention effects. This task coincides with inferring causal effects, a well-established area in Statistics (e.g., [5, 8, 10, 11, 13, 18, 24, 25, 26] and [31]). We emphasize that, in our application, it is exactly the intervention or causal effect that is of interest, rather than a regression-type effect of association. If we can estimate the intervention effects from observational data, we can score each gene according to its potential to have an intervention (knock-down) effect on the riboflavin production rate, and the most promising candidate genes can be tested afterward in biological experiments.

Pearl ([25], page 285) formulates the distinction between associational and causal concepts as follows: "an associational concept is any relationship that can be defined in terms of a joint distribution of observed variables, and a causal concept is any relationship that cannot be defined from the distribution alone.... Every claim invoking causal concepts must be traced to some premises that invoke such concepts; it cannot be inferred or derived from statistical associations alone." Thus, in order to obtain causal statements from observational data, one needs to make additional assumptions. One possibility is to assume that the data were generated by a directed acyclic graph (DAG) which is *known* beforehand. DAGs describe causal concepts, since they code potential causal relationships between variables: the existence of a directed edge $x \to y$ means that $x$ *may* have a direct causal effect on $y$, and the absence of a directed edge $x \to y$ means that $x$ *cannot* have a direct causal effect on $y$ (see Remark 2.3 for a definition of direct causal effect).

Given a set of conditional dependencies from observational data and a corresponding DAG model, one can compute causal effects using intervention calculus (e.g., [24] and [25]). In this paper, we consider the problem of inferring causal information from observational data, under the assumption that the data were generated by an *unknown* DAG. This is a more realistic assumption, since, in many practical problems, one does not know the DAG. In this scenario, the causal effect is typically not defined uniquely, and that is not surprising, given the description of causality by Pearl [25] above.

A DAG is typically not identifiable from observational data, because conditional dependencies only determine the skeleton and the so-called $v$-structures of the graph. The skeleton and $v$-structures determine an equivalence class of DAGs that all correspond to the same probability distribution. This equivalence class can be described by a completed partially directed acyclic graph (CPDAG) (see Section 2.1).

The existence of the equivalence class opens the way to the following strategy. Suppose that we are interested in the causal effects of a covariate $X_i$ on a response $Y$. We are given the joint distribution of $X_1, \ldots, X_p, Y$, and use this to find the equivalence class of DAGs that correspond to this distribution. Assume that this equivalence class contains $m$ different DAGs. For each DAG $G_j$ in this class, we can apply intervention calculus to obtain the causal effect $\theta_{ij}$ of $X_i$ on $Y$. We can summarize this information for $i = 1, \ldots, p$ and $j = 1, \ldots, m$ in a $p \times m$ matrix $\Theta$, where each row corresponds to a covariate and each column corresponds to a DAG in the equivalence class. Since the ordering of the DAGs in the equivalence class is arbitrary, the columns of this matrix can be permuted in any order. It is our goal to estimate this matrix $\Theta$. A slightly less ambitious goal is to estimate the multisets $\Theta_i = \{\theta_{ij}\}_{j \in \{1,\ldots,m\}}$, $i = 1, \ldots, p$, containing the possible causal effects of covariate $X_i$ on $Y$ (see Section 3.2 for the definition of a multiset). Note that $\Theta$ contains slightly more information than $\Theta_i$, $i = 1, \ldots, p$, since the columns of $\Theta$ tell us which possible causal effects originate from the same DAG, while this information is lost in the multisets $\Theta_i$, $i = 1, \ldots, p$.

In special cases, all values $\theta_{ij}$, $j = 1, \ldots, m$, in $\Theta_i$ may be identical, so that the causal effect of $X_i$ on $Y$ is uniquely determined. But, even if $\Theta_i$ contains distinct values, it still contains useful causal information. For example, if $\theta_{ij} \neq 0$ for all $j = 1, \ldots, m$, then $X_i$ must have a causal effect on $Y$ (positive or negative). Similarly, if $\theta_{ij} > 0$ for all $j = 1, \ldots, m$, then $X_i$ must have a positive causal effect on $Y$. Finally, the minimum absolute value $\min_j |\theta_{ij}|$ is a lower bound on the size of the causal effect of $X_i$ on $Y$. We use this bound to determine variable importance.

There is a large existing literature on estimating the equivalence class of DAGs (e.g., [2, 3, 4, 12, 14, 30, 31] and [33]), and there is also a large literature on estimating causal effects when a DAG is given (e.g., [18, 19, 23, 24] and [25]). Our new approach combines these two parts in order to estimate the multisets of possible causal effects $\Theta_i$, $i = 1, \ldots, p$. We use these multisets to determine bounds for causal effects and causal importance of variables. We also show that the distinct values of $\Theta_i$ can be estimated by a new algorithm that uses only *local* information of the estimated CPDAG, thus allowing for efficient computation in very large problems, and we prove that this method is asymptotically consistent in sparse high-dimensional settings.

The outline of this paper is as follows. In Section 2, we introduce terminology for graphs and intervention calculus. Sections 3 and 4 discuss our proposed methodology to estimate the multisets of possible causal effects $\Theta_i$, $i = 1, \ldots, p$. Section 3 discusses so-called population versions of the algorithms that can be used if all conditional dependencies are known exactly. Section 4 discusses sample versions of the algorithms that can be used if the conditional dependencies are estimated from data. In Section 5, we prove asymptotic consistency of our methods in high-dimensional settings with certain sparsity and regularity assumptions. In Section 6, we evaluate our methods in a simulation study, and apply them to the riboflavin data set. Finally, Section 7 contains a brief discussion, Section 8 contains collected proofs and the Appendix contains a description of possible modifications of the algorithms.

## 2. Graph terminology and intervention calculus.

2.1. *Graphs.* Let $G = (V, E)$ be a graph consisting of vertices $V$ and a set of edges $E \subseteq V \times V$. In our context, the vertices represent random variables $X_1, \ldots, X_p$ and $Y$, and the edges represent relationships between pairs of these variables.

An edge between two vertices, say $X_i$ and $X_j$, is *directed* if the edge has an arrowhead, that is, $X_i \leftarrow X_j$ or $X_i \rightarrow X_j$. An edge between $X_i$ and $X_j$ is *undirected* if it has no arrowhead, that is, $X_i - X_j$. A *directed graph* is a graph in which all edges are directed. An *undirected graph* is a graph in which all edges are undirected. A *partially directed graph* may contain both directed and undirected edges. The *skeleton* of a (partially) directed graph $G$ is the undirected graph that is obtained from $G$ by removing all arrowheads.

Two vertices $X_i$ and $X_j$ are *adjacent* if there is a directed or undirected edge between them. The *adjacency set* of a vertex $X_i$, denoted by $\mathrm{adj}_i(G)$, is the collection of all vertices that are adjacent to $X_i$ in $G$. A *path* is any unbroken nonintersecting route that can be traced along the edges of the skeleton of the graph. A *directed path* is a path along directed edges that follows the direction of the arrows. A (*directed*) *cycle* is a (directed) path that starts and ends at the same vertex. A graph that contains no directed cycles is called *acyclic*. A graph that is both directed and acyclic is called a *directed acyclic graph* (*DAG*). A *v-structure* in a graph $G$ is an ordered triple of vertices, say $(X_i, X_j, X_k)$, such that $G$ contains directed edges $X_i \rightarrow X_j$ and $X_j \leftarrow X_k$, and $X_i$ and $X_k$ are not adjacent in $G$. The vertex $X_j$ is then called a *collider* in this *v*-structure.

Consider a partially directed graph $G$. Vertex $X_j$ is said to be a *parent* of $X_i$ in $G$ if there is a directed edge $X_j \rightarrow X_i$. The set of all parents of $X_i$ in $G$ is denoted by $\mathrm{pa}_i(G)$. Vertex $X_j$ is said to be a *sibling* of $X_i$ in $G$ if there is an undirected edge $X_i - X_j$. The set of all siblings of $X_i$ in $G$ is denoted by $\mathrm{sib}_i(G)$. For any subset $S$ of $\mathrm{sib}_i(G)$, we let $G_{S \rightarrow i}$ denote the graph that is obtained by changing all undirected edges $X_j - X_i$ with $X_j \in S$ into directed edges $X_i \leftarrow X_j$, and all undirected edges $X_j - X_i$ with $X_j \in \mathrm{sib}_i(G) \setminus S$ into directed edges $X_i \rightarrow X_j$. If the graph $G$ is clear from the context, we write $\mathrm{pa}_i$ and $\mathrm{sib}_i$ instead of $\mathrm{pa}_i(G)$ and $\mathrm{sib}_i(G)$.

A DAG encodes conditional independence relationships via the notion of *d-separation* ([24], Definition 1.2.3, page 16). A distribution $P$ is said to be *faithful* to a graph $G$ if the conditional independence relationships of $P$ are exactly the same as those encoded by $G$ via *d*-separation. In general, the same set of conditional independence relationships can be described by several DAGs. These DAGs form an *equivalence class*, consisting of DAGs with the same skeleton and the same $v$-structures [33]. Such an equivalence class can be uniquely described by a *completed partially directed acyclic graph* (CPDAG) [2]. This is a partially directed graph with the same skeleton as the graphs in the equivalence class in which the edges are directed as follows: (i) the directed edges represent arrows that are common to all DAGs in the equivalence class, and (ii) the undirected edges correspond to edges that are directed one way in some DAGs and the other way in other DAGs in the equivalence class. We say that a partially directed graph $G$ is *extendable* to a DAG if its undirected edges can be directed without creating directed cycles or additional $v$-structures.

A CPDAG can be estimated in various ways, including the PC-algorithm [31], search and score methods (cf. [2, 3, 4] and [33]) and Bayesian methods (cf. [12] and [30]). In this paper, we will use the PC-algorithm, since this algorithm is computationally feasible and asymptotically consistent in sparse high-dimensional settings [14]. We refer to [28] and [35] for a discussion about pointwise versus uniform consistency of the PC-algorithm.

2.2. *Intervention calculus.* We now give a brief Introduction to intervention calculus, mostly based on [24] and [25]. We consider $p+1$ variables $X_1, \ldots, X_p, Y$ (sometimes also referred to as $X_1, \ldots, X_{p+1}$).

Any distribution that is generated from a DAG with independent error terms is called Markovian, with respect to the DAG. Any Markovian distribution can be factorized as

$$f(x_1, \ldots, x_{p+1}) = \prod_{j=1}^{p+1} f(x_j | \mathrm{pa}_j)$$

see [25], Theorem 3.1, page 297; see also [17], Section 3.2.2, for a formulation in terms of directed local or global Markov properties.

In order to represent the effect of an intervention on a set of variables, [16] and [23] introduced so-called *do* or *set* operators. In particular, they used expressions of the form $f(y|\mathrm{do}(X_i = x'_i))$ or $f(y|\mathrm{set}(X_i = x'_i))$ to denote the distribution of $Y$ that would occur if treatment condition $X_i = x'_i$ was enforced uniformly over the population via some intervention. For a Markovian model, the distribution generated by an intervention $\mathrm{do}(X_i = x'_i)$ on the set of variables $X_1, \ldots, X_{p+1}$ is given by the following truncated factorization formula:

$$(1) \quad f(x_1, \ldots, x_{p+1} | \mathrm{do}(X_i = x'_i)) = \begin{cases} \prod_{j=1, j \neq i}^{p+1} f(x_j | \mathrm{pa}_j)|_{x_i = x'_i}, & \text{if } x_i = x'_i, \\ 0, & \text{otherwise,} \end{cases}$$

where $f(x_j|\mathrm{pa}_j)$ are the pre-intervention conditional distributions ([25], Corollary 3.1, page 297). Note that this formula uses the DAG structure (determining the sets $\mathrm{pa}_j$) to write the interventional distribution on the left-hand side in terms of pre-intervention conditional distributions on the right-hand side.

The distribution of $Y = X_{p+1}$ after an intervention $\mathrm{do}(X_i = x'_i)$ can be found by integrating out $x_1, \ldots, x_p$ in (1). It can be shown that this simplifies to the following:

$$
(2) \quad f(y|\mathrm{do}(X_i = x'_i)) = \begin{cases} f(y), & \text{if } Y \in \mathrm{pa}_i, \\ \int f(y|x'_i, \mathrm{pa}_i) f(\mathrm{pa}_i) \, d\mathrm{pa}_i, & \text{if } Y \notin \mathrm{pa}_i, \end{cases}
$$

where $f(\cdot)$ and $f(\cdot|x'_i, \mathrm{pa}_i)$ represent pre-intervention distributions ([24], Theorem 3.2.2, page 73). Note that the expression in (2) for $Y \notin \mathrm{pa}_i$ is a special case of so-called *back-door adjustment* ([24], Theorem 3.3.2, page 79) since $\mathrm{pa}_i$ satisfies the *back-door criterion* relative to $(X_i, Y)$ if $Y \notin \mathrm{pa}_i$ ([24], Definition 3.3.1, page 79).

It is common ([24], page 70) to summarize the distribution generated by an intervention by its mean

$$
E(Y|\mathrm{do}(X_i = x'_i)) = \begin{cases} E(Y), & \text{if } Y \in \mathrm{pa}_i, \\ \int E(Y|x'_i, \mathrm{pa}_i) f(\mathrm{pa}_i) \, d\mathrm{pa}_i, & \text{if } Y \notin \mathrm{pa}_i, \end{cases}
$$

and we can then define the *causal effect* of $\mathrm{do}(X_i = x'_i)$ on $Y$ by

$$
(3) \qquad \frac{\partial}{\partial x} E(Y|\mathrm{do}(X_i = x))|_{x = x'_i}.
$$

In the remainder of the paper, we consider the case that $X_1, \ldots, X_p, Y$ are jointly Gaussian, and we are interested in the causal effect of $X_i$ on $Y$ for $i = 1, \ldots, p$. In this case, it is very simple to compute the causal effects as defined in (3), since Gaussianity implies that $E(Y|x'_i, \mathrm{pa}_i)$ is linear in $x'_i$ and $\mathrm{pa}_i$

$$
E(Y|x'_i, \mathrm{pa}_i) = \gamma_0 + \gamma_i x'_i + \gamma_{\mathrm{pa}_i}^T \mathrm{pa}_i
$$

for some values $\gamma_0, \gamma_i \in \mathbb{R}$ and $\gamma_{\mathrm{pa}_i} \in \mathbb{R}^{|\mathrm{pa}_i|}$, where $|\mathrm{pa}_i|$ is the cardinality of the set $\mathrm{pa}_i$. Hence,

$$
\int E(Y|x'_i, \mathrm{pa}_i) f(\mathrm{pa}_i) \, d\mathrm{pa}_i = \gamma_i x'_i + \int \gamma_{\mathrm{pa}_i}^T \mathrm{pa}_i f(\mathrm{pa}_i) \, d\mathrm{pa}_i
$$

is linear in $x'_i$. Combining this with (3), it follows that the causal effect of $X_i$ on $Y$ with $Y \notin \mathrm{pa}_i$ is given by $\gamma_i$, which is simply the regression coefficient of $X_i$ in the regression of $Y$ on $X_i$ and $\mathrm{pa}_i$. In general, the causal effect of $X_i$ on $Y$ as defined in (3) is given by $\beta_{i|\mathrm{pa}_i}$, where, for any set $S \subseteq \{X_1, \ldots, X_p, Y\} \setminus \{X_i\}$,

$$
(4) \qquad \beta_{i|S} = \begin{cases} 0, & \text{if } Y \in S, \\ \text{coefficient of } X_i \text{ in } Y \sim X_i + S, & \text{if } Y \notin S, \end{cases}
$$

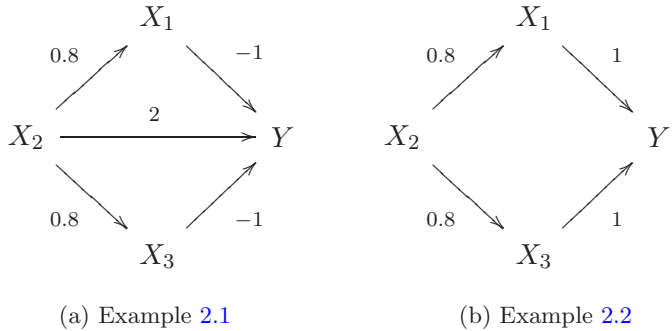

(a) Example 2.1        (b) Example 2.2

FIG. 1. *Graphical representation of the models used in Examples 2.1 and 2.2.*

and $Y \sim X_i + S$ is shorthand for the linear regression of $Y$ on $X_i$ and $S$. Hence, in the Gaussian case, the causal effect does not depend on the value of $x'_i$, and can be interpreted as

$$E(Y|\text{do}(X_i = x'_i + 1)) - E(Y|\text{do}(X_i = x'_i))$$

for any value of $x'_i$.

2.3. *Intervention calculus versus association.* In the previous section, we discussed that, for jointly Gaussian variables, intervention effects can be computed using linear regression. We emphasize, however, that intervention calculus and multiple regression analysis generally give different results, since the set of variables that is controlled for is different. We illustrate this difference using two examples. In Example 2.1, the variable that appears to be most important in the regression analysis is least important in the causal analysis. Example 2.2 shows that the opposite is also possible. The variable that has no importance in the regression analysis is most important in the causal analysis. Throughout, we will use $\beta$ to denote the regression parameters, and $\theta$ to denote the intervention effects.

EXAMPLE 2.1. Consider the following model [see Figure 1(a)]: $X_2 = \varepsilon_2$, $X_1 = 0.8X_2 + \varepsilon_1$, $X_3 = 0.8X_2 + \varepsilon_3$ and

$$Y = -X_1 + 2X_2 - X_3 + \varepsilon,$$

where $\varepsilon_1, \varepsilon_2, \varepsilon_3$ and $\varepsilon$ are mutually independent normal random variables with mean zero and variances $\sigma_1^2 = 0.36$, $\sigma_2^2 = 1$, $\sigma_3^2 = 0.36$ and $\sigma^2 = 1$. Note that $X_1$, $X_2$ and $X_3$ all have variance 1, so that we can meaningfully compare their regression coefficients or causal effects.

First suppose that we apply multiple linear regression $Y = \alpha + \beta_1 X_1 + \beta_2 X_2 + \beta_3 X_3 + \varepsilon$. Then the regression coefficients are $\beta_1 = -1$, $\beta_2 = 2$ and $\beta_3 = -1$. Looking at the sizes of the effects, variable $X_2$ is most important in the regression analysis.

Next, we apply intervention calculus. Let $\theta = (\theta_1, \theta_2, \theta_3)$, where $\theta_i$ represents the causal effect of $X_i$ on $Y$. Since $\text{pa}_1 = \{X_2\}$, $\text{pa}_2 = \varnothing$ and $\text{pa}_3 = \{X_2\}$, we have $\theta_1 = \beta_{1|X_2} = -1$, $\theta_2 = \beta_{2|\varnothing} = 0.4$ and $\theta_3 = \beta_{3|X_2} = -1$. We see that $\theta_1 = \beta_1$ and $\theta_3 = \beta_3$, but that $\theta_2 \neq \beta_2$. Considering the sizes of the causal effects, variable $X_2$ is least important in the causal analysis.

EXAMPLE 2.2. Let $X_1$, $X_2$ and $X_3$ be as in Example 2.1, and let

$$Y = X_1 + X_3 + \varepsilon$$

[Figure 1(b)]. Applying multiple linear regression $Y = \alpha + \beta_1 X_1 + \beta_2 X_2 + \beta_3 X_3 + \varepsilon$, the regression coefficients are $\beta_1 = 1$, $\beta_2 = 0$ and $\beta_3 = 1$. Looking at the sizes of the effects, variable $X_2$ is least important.

On the other hand, if we consider intervention calculus, we get $\theta_1 = \beta_{1|X_2} = 1$, $\theta_2 = \beta_{2|\varnothing} = 1.6$ and $\theta_3 = \beta_{3|X_2} = 1$. We again see that $\theta_1 = \beta_1$ and $\theta_3 = \beta_3$, but that $\theta_2 \neq \beta_2$. Considering the sizes of the causal effects, variable $X_2$ is now most important.

REMARK 2.3. In Examples 2.1 and 2.2, $Y$ is not a parent of any of the $X$'s. For such DAGs, we can formulate the distinction between intervention calculus and multiple regression as follows. The causal effect $\theta_i$ measures the *total* effect of variable $X_i$ on the response $Y$ (i.e., the sensitivity of $Y$ to interventional changes in $X_i$). On the other hand, the regression parameter $\beta_i$ measures the *direct* effect of $X_i$ on $Y$ (i.e., the sensitivity of $Y$ to interventional changes in $X_i$ when all other variables in the model are held fixed). For a precise definition of direct effect (see [24], pages 126–127).

**3. Population versions of the algorithms.** The intervention calculus discussed in Section 2.2 assumes that the DAG that generates the distribution of $X_1, \ldots, X_p, Y$ is known. We now present our new methodology for determining causal effects when the DAG is *unknown*. First, in Section 3.1, we state our assumptions. In Section 3.2, we discuss our methods, assuming that all conditional dependencies are known exactly (hence the terminology *population versions*). Section 4 will treat *sample versions* of the algorithms, that is, versions of the algorithms that can be used if the conditional dependencies are estimated from the data. We split the exposition in these two parts, since this allows us to separate the main ideas of the methods (Section 3) from the extra complications that arise from working with estimated conditional dependencies (Section 4).

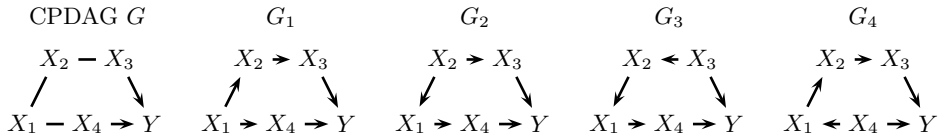

Fig. 2. *A CPDAG $G$ with the DAGs $G_1, \ldots, G_4$ that are in its equivalence class.*

3.1. *Assumptions.* We make the following assumptions:

(A) The distribution of $(X_1, \ldots, X_p, Y)$ is multivariate normal. Moreover, it is Markovian and faithful to the true (unknown) causal DAG.
(B) $X_1, \ldots, X_p$ have equal variance.

The Gaussianity assumption in (A) implies that $E(Y|S)$ is linear for any $S \subseteq \{X_1, \ldots, X_p\}$, so that the causal effects can be easily computed (see Section 2.2). Moreover, it allows us to equate conditional independence with zero partial correlation. This is useful in the PC-algorithm [31], which we employ to find the equivalence class of DAGs. Faithfulness is also used in the PC-algorithm. It makes it possible to move hierarchically from marginal or low-order partial correlations to higher orders, yielding a tremendous computational advantage if $p$ is large. Both normality and faithfulness are used to prove consistency of our methods (see Section 5). Assumption (B) is made for convenience, so that we can easily compare the causal effects of different variables.

3.2. *The algorithms.* In the population versions of the algorithms, we assume that all conditional dependencies are known exactly. In this case, the population version of the PC-algorithm (see [14] and [31] for a detailed description) yields the correct CPDAG.

Based on this CPDAG, we can compute the sets of possible causal effects. Before describing the algorithms to do this, we note that the output of the algorithms consists of *multisets*. A multiset is similar to a set, with the only difference that in a multiset the multiplicity of elements matters. Thus, the multisets $\{a, b\}$ and $\{b, a\}$ are equal, just as the sets $\{a, b\}$ and $\{b, a\}$, since the order of the elements does not matter. But the multisets $\{a, a\}$ and $\{a\}$ are not equal, while the sets $\{a, a\}$ and $\{a\}$ are.

The basic idea of our method is given in pseudocode in Algorithm 1. We illustrate this algorithm by computing $\Theta_1$, the set of possible causal effects of $X_1$ on $Y$, for the CPDAG $G$ in Figure 2. First, we list all DAGs in the equivalence class of $G$. Note that $G$ contains the 3 undirected edges $X_1 - X_2$, $X_1 - X_4$ and $X_2 - X_3$. There are 8 possible ways to direct these edges, but some of these lead to graphs that are not in the equivalence class of $G$. For example, the configuration $X_1 \to X_2$, $X_1 \to X_4$ and $X_2 \leftarrow X_3$ is invalid, since this creates a new $v$-structure $X_1 \to X_2 \leftarrow X_3$ and that is incompatible with the equivalence class represented by $G$ (see Section 2.1). Excluding such invalid configurations leaves four DAGs in the equivalence class of $G$ (see $G_1, \ldots, G_4$ in Figure 2). Next, for each $j = 1, \ldots, 4$, we compute the causal effect $\theta_{1j}$ of $X_1$ on $Y$, assuming the data were generated from DAG $G_j$. Using (4) and assumption (A) of Section 3.1, this yields

$$
\begin{aligned}
\Theta_1 &= \{\theta_{11}, \theta_{12}, \theta_{13}, \theta_{14}\} = \{\beta_{1|\mathrm{pa}_1(G_1)}, \beta_{1|\mathrm{pa}_1(G_2)}, \beta_{1|\mathrm{pa}_1(G_3)}, \beta_{1|\mathrm{pa}_1(G_4)}\} \\
&= \{\beta_{1|\varnothing}, \beta_{1|X_2}, \beta_{1|X_2}, \beta_{1|X_4}\}.
\end{aligned}
\tag{5}
$$

Note that the parental sets of $X_1$ in the four DAGs in the equivalence class of $G$ play a crucial role in determining the possible causal effects of $X_1$ on $Y$. In particular, since $\mathrm{pa}_1(G_1) = \varnothing$, $\mathrm{pa}_1(G_2) = \mathrm{pa}_1(G_3) = \{X_2\}$, and $\mathrm{pa}_1(G_4) = \{X_4\}$, the multiset $\Theta_1$ contains $\beta_{1|\varnothing}$ with multiplicity 1, $\beta_{1|X_2}$ with multiplicity 2, and $\beta_{1|X_4}$ with multiplicity 1.

The basic Algorithm 1 works well if the number of covariates is small, say less than 10 or so. But, if the number of covariates increases, it quickly becomes infeasible to compute all DAGs in the equivalence class. We therefore developed a localized algorithm which is much faster. In order to explain this local algorithm, we first discuss a variation on the basic algorithm, given in pseudocode in Algorithm 2.

Algorithm 2 is based on the idea that, for the computation of $\Theta_1$, the parents of $X_1$ in the different DAGs in the equivalence class are of key importance. Therefore, we first consider the CPDAG $G$ and determine all possible parental sets of $X_1$; that is, we take all sets $\mathrm{pa}_1(G) \cup S$ where $S \subseteq \mathrm{sib}_1(G)$. In Figure 2, $\mathrm{pa}_1(G) = \varnothing$ and $\mathrm{sib}_1(G) = \{X_2, X_4\}$, so that the possible parental sets of $X_1$ are $\varnothing$, $\{X_2\}$, $\{X_4\}$ and $\{X_2, X_4\}$. These sets $S$ determine the direction of the edges between $X_1$ and the vertices in $\mathrm{sib}_1(G)$. All edges between $X_1$ and vertices in $S$ must be directed toward $X_1$, and all edges between $X_1$ and vertices in $\mathrm{sib}_1(G) \setminus S$ must be directed away from $X_1$, exactly as in $G_{S \to 1}$ (see Section 2.1). For each set $S$, we then determine the number of DAGs $m_S$ to which $G_{S \to 1}$ is extendable. As illustration, we compute $m_S$ for $S = \{X_2\}$ and $S = \{X_4\}$. First, note that $S = \{X_2\}$ implies

---

**Algorithm 1**: Basic algorithm

  **Input**: CPDAG $G$, conditional dependencies of $X_1, \ldots, X_p, Y$
  **Output**: Matrix $\Theta$ of possible causal effects
**1** Determine all DAGs $G_1, \ldots, G_m$ in the equivalence class of $G$
**2 for** $j = 1$ *to* $m$ **do**
**3**     **for** $i = 1$ *to* $p$ **do**
**4**         $\theta_{ij} = \beta_{i|\mathrm{pa}_i(G_j)}$ [see (4)]
**5**     **end**
**6 end**

**Algorithm 2**: Variation on Algorithm 1 (for instructive purposes)

**Input**: CPDAG G, conditional dependencies of $X_1, \ldots, X_p, Y$
**Output**: Multisets $\Theta_1, \ldots, \Theta_p$ of possible causal effects

**1 for** $i = 1$ *to* $p$ **do**
**2**     $\Theta_i = \varnothing$
**3**     **for each** *subset* $S$ of $\mathrm{sib}_i(G)$ **do**
**4**         $m_S$ = number of DAGs to which $G_{S \to i}$ is extendable
**5**         add $m_S$ copies of $\beta_{i|\mathrm{pa}_i(G) \cup S}$ to $\Theta_i$
**6**     **end**
**7 end**

that $X_1 \leftarrow X_2$ and $X_1 \to X_4$, since $X_2$ is a parent of $X_1$ and $X_4$ is not. The undirected edge $X_2 - X_3$ in $G_{S \to 1}$ can then be directed both ways without creating a new $v$-structure or a directed cycle. Hence, for $S = \{X_2\}$ we have $m_S = 2$. On the other hand, $S = \{X_4\}$ implies $X_1 \to X_2$ and $X_1 \leftarrow X_4$. In this case, the undirected edge $X_2 - X_3$ in $G_{S \to 1}$ must be directed toward $X_3$, since otherwise a new $v$-structure $X_1 \to X_2 \leftarrow X_3$ is created. Hence, for $S = \{X_4\}$ we have $m_S = 1$. Using the same reasoning for $S = \varnothing$ and $S = \{X_2, X_4\}$, one can easily check that the multiplicities corresponding to $S = \varnothing, \{X_2\}, \{X_4\}, \{X_2, X_4\}$ are $m_S = 1, 2, 1, 0$. Finally, we form the multiset $\Theta_1$ by taking the elements $\beta_{1|\mathrm{pa}_1(G) \cup S}$ with multiplicities $m_S$, for all $S \subseteq \mathrm{sib}_1(G)$ (where elements with multiplicity zero are omitted). Thus, in Figure 2, we obtain $\Theta_1 = \{\beta_{1|\varnothing}, \beta_{1|X_2}, \beta_{1|X_2}, \beta_{1|X_4}\}$.

From this construction, it is clear that Algorithm 2 gives the same output as Algorithm 1 (with the only difference that Algorithm 2 does not yield the column structure of $\Theta$, telling us which causal effects originate from the same DAG). Note that Algorithm 2 is not faster than Algorithm 1. The new bottleneck is the computation of the multiplicities $m_S$, which again quickly becomes infeasible if the number of covariates increases. We therefore do not recommend using this algorithm in practice. However, we can slightly modify Algorithm 2 to obtain a fast localized algorithm, given in pseudocode in Algorithm 3.

The difference between Algorithms 2 and 3 is that Algorithm 3 replaces the computation of $m_S$ by a much simpler step which only checks if $G_{S \to i}$ is *locally valid*, meaning that $G_{S \to i}$ does not contain an additional $v$-structure with $X_i$ as collider. In the example in Figure 2, $G_{S \to 1}$ is locally valid for $S = \varnothing, \{X_2\}$ and $\{X_4\}$, and it is not locally valid for $S = \{X_2, X_4\}$. We then form a new multiset $\Theta_1^L$ by taking all elements $\beta_{1|\mathrm{pa}_1(G) \cup S}$ for which $G_{S \to 1}$ is locally valid. In the example, this results in $\Theta_1^L = \{\beta_{1|\varnothing}, \beta_{1|X_2}, \beta_{1|X_4}\}$.

Note that, for the CPDAG in Figure 2, the sets of distinct values in $\Theta_1^L$ and $\Theta_1$ are the same, but the multiplicities are different. It turns out that this holds in general. To show this, we need the following lemma.

**Algorithm 3**: Local algorithm

**Input**: CPDAG G, conditional dependencies of $X_1, \ldots, X_p, Y$
**Output**: Multisets $\Theta_i^L$, $i = 1, \ldots, p$

**1 for** $i = 1$ *to* $p$ **do**
**2**   $\Theta_i^L = \varnothing$
**3**   **for each** *subset* $S$ of $\text{sib}_i(G)$ **do**
**4**      **if** $G_{S \to i}$ *is locally valid (i.e., has no new v-structure with collider* $X_i$*)* **then**
**5**         add $\beta_{i|\text{pa}_i(G) \cup S}$ to $\Theta_i^L$
**6**      **end**
**7**   **end**
**8 end**

LEMMA 3.1. *Let $S \subseteq \text{sib}_i(G)$. Then $G_{S \to i}$ is locally valid if and only if there is a DAG $G_j$ in the equivalence class of $G$ such that $\text{pa}_i(G_j) = \text{pa}_i(G) \cup S$.*

One direction of this lemma is trivial. If there is a DAG $G_j$ in the equivalence class of $G$ with $\text{pa}_i(G_j) = \text{pa}_i(G) \cup S$, then by definition $G_j$ is locally valid and, hence, $G_{S \to i}$ must be locally valid. Surprisingly, the other direction also holds, as proved in Section 8.

Lemma 3.1 directly leads to the following result.

THEOREM 3.2. $\Theta_i$ *and* $\Theta_i^L$ *are equal when they are interpreted as sets:*

$$\Theta_i \stackrel{\text{set}}{=} \Theta_i^L, \qquad i = 1, \ldots, p.$$

Theorem 3.2 implies that the only information we lose by using the local Algorithm 3 is the multiplicity of the values. The sets of distinct values in $\Theta_i^L$ and $\Theta_i$ are exactly the same. Implications of this result are that, for example, the range of possible causal effects or the minimum absolute value of the possible causal effects can be obtained via the local Algorithm 3.

REMARK 3.3. Note that the multiplicities of elements in $\Theta_i$ and $\Theta_i^L$ have different meanings. The multiplicity of an element $\theta$ in $\Theta_i$ corresponds to the *number of DAGs* in the equivalence class for which the causal effect of $X_i$ on $Y$ equals $\theta$. On the other hand, the multiplicity of an element $\theta'$ in $\Theta_i^L$ corresponds to the *number of subsets $S$* in the local Algorithm 3 that yield causal effect $\theta'$. The cardinality of $\Theta_i^L$ is always smaller or equal to the cardinality of $\Theta_i$, since each set $S$ in Algorithm 3 corresponds to at least one DAG in the equivalence class (Lemma 3.1).

**4. Sample versions of the algorithms.** Assume that we have a sample consisting of $n$ i.i.d. copies of $(X_1,\ldots,X_p,Y) = (X_1,\ldots,X_{p+1})$. We then obtain sample versions of the algorithms by using the estimated conditional dependencies of $X_1,\ldots,X_p,Y$ as input. In the Gaussian case, we use estimated partial correlations $\hat{\rho}_{nij|S}$ between $X_i$ and $X_j$ given some set of other variables $S$. We then use the sample version of the PC-algorithm to estimate the corresponding CPDAG $G$ [14] and [31]. This involves multiple testing for $Z$-transformed partial correlations

$$\hat{Z}_{nij|S} = \frac{1}{2}\log\left(\frac{1+\hat{\rho}_{nij|S}}{1-\hat{\rho}_{nij|S}}\right).$$

Since $\hat{Z}_{nij|S}$ has a $N(0,(n-|S|-3)^{-1})$ distribution if $\rho_{ij|S}=0$, we conclude that $\rho_{ij|S} \neq 0$ if

$$|\hat{Z}_{nij|S}|\sqrt{n-|S|-3} > \Phi^{-1}(1-\alpha/2),$$

where $\Phi$ is the standard normal distribution function and $0 < \alpha < 1$ is a tuning parameter.

Next, we use the estimated CPDAG $\hat{G}(\alpha)$ to estimate the multisets of possible causal effects, by using sample versions of (4) (i.e., we use the least squares estimated regression coefficients). This procedure has been implemented in the R-package pcalg [15] (in the meantime, code is available from the authors). We denote the estimated multisets by

$\hat{\Theta}_{ni}(\alpha)$ for the sample version of the basic Algorithm 1,

$\hat{\Theta}_{ni}^L(\alpha)$ for the sample version of the local Algorithm 3

for $i=1,\ldots,p$, where we emphasize the dependence of the estimates on the tuning parameter $\alpha$. Possible modifications of Algorithms 1 and 3 that can be beneficial in the sample versions of the algorithms are discussed in the Appendix.

4.1. *Tuning of the PC-algorithm.* The tuning parameter $\alpha$ in the PC-algorithm can be chosen via a Bayesian Information Criterion (BIC). First, for a given choice of $\alpha$, we compute the estimated CPDAG $\hat{G}(\alpha)$. Next, we find a DAG $\hat{G}'(\alpha)$ that is in the equivalence class described by $\hat{G}(\alpha)$. Based on $\hat{G}'(\alpha)$, we then compute the maximum likelihood estimators $\hat{\Sigma}_{\text{MLE},\hat{G}'(\alpha)}$ and $\hat{\mu}_{\text{MLE}}$ for the covariance matrix and mean vector of the Gaussian distribution of $X_1,\ldots,X_{p+1}$ (cf. [19]). Finally, we choose $\alpha$ to minimize

$$-2\ell(\hat{\Sigma}_{\text{MLE},\hat{G}'(\alpha)},\hat{\mu}_{\text{MLE}}) + \log n\left(\sum_{i\leq j}1_{(\hat{\Sigma}_{\text{MLE},\hat{G}'(\alpha)})_{ij}\neq 0} + p+1\right),$$

where $\ell(\cdot)$ denotes the log-likelihood of a $(p+1)$-dimensional multivariate Gaussian distribution. We point out that the behavior of BIC is still unknown in the high-dimensional setting, where the dimensionality $p$ may be much larger than the sample size $n$.

Another approach to tune the PC-algorithm, is to choose $\alpha$ relatively large, so that the resulting graph contains a large number of edges. We then investigate which edges (directed or undirected) are stable under a subsampling procedure, where stability is measured in terms of the relative frequency of occurrence of (directed or undirected) edges under the sub-sampling scheme. An edge is kept if the corresponding subsampling frequency is larger than a certain cut-off. Surprisingly, this cut-off can be determined via controlling a multiple testing error rate. Details of such a generic procedure are described in [22].

4.2. *Incoherences with sample versions.* Two types of incoherences may occur in the sample version of the PC-algorithm (but the probability of these incoherences converges to zero as the sample size $n$ goes to infinity).

First, the sample version of the PC-algorithm may produce conflicting $v$-structures. For example, the algorithm can produce $v$-structures $X_1 \to X_2 \leftarrow X_3$ and $X_2 \to X_3 \leftarrow X_4$, giving conflicting information about the direction of the edge $X_2 - X_3$. In such cases, the algorithm overwrites the $v$-structures in the order in which they were tested. Hence, the resulting structure depends on the order in which the independence tests are performed. Since we usually do not prefer one order of tests over another, we simply choose the structure that arises by the ordering of the variables.

Second, the sample version of the PC-algorithm may produce invalid CPDAGs (i.e., CPDAGs that are not extendable). For example, the algorithm may yield a graph with undirected edges $X_1 - X_2$, $X_2 - X_3$, $X_3 - X_4$ and $X_4 - X_1$. This is not a valid CPDAG, since it is impossible to direct its edges without creating a directed cycle or a $v$-structure. In other words, this graph does not describe an equivalence class of DAGs. While such an invalid CPDAG does not cause problems in the local Algorithm 3, it is problematic in the basic Algorithm 1, since in the latter algorithm the CPDAG has to be extended in order to find all DAGs in the equivalence class. In Algorithm 1, we solve this problem by modifying the estimated CPDAG in the following way. First, we search for conflicting $v$-structures, and we try to rearrange them until we get an extendable CPDAG. If this is not possible, we destroy as few $v$-structures as possible to obtain an extendable CPDAG. If this is also not possible, we randomly draw a DAG on the estimated skeleton and work with the CPDAG that corresponds to this DAG.

**5. Asymptotic consistency.** In this section, we prove asymptotic consistency of our methods in high-dimensional settings (i.e., in situations where the number of covariates $p$ can be much larger than the sample size $n$). We consider a framework where the model depends on $n$. We use $p_n$ to denote the number of covariates, $G_n$ to denote the CPDAG, and $P_n$ to denote the distribution of $(X_{n1}, \ldots, X_{np_n}, Y_n) = (X_{n1}, \ldots, X_{np_n}, X_{n,p_n+1})$. We assume that the data consist of $n$ i.i.d. copies of $(X_{n1}, \ldots, X_{n,p_n+1}) \sim P_n$. Regarding $P_n$, we make assumption (A) of Section 3.1. Additionally, we assume the following:

(C) The number of covariates $p_n = O(n^a)$ for some $0 \leq a < \infty$.
(D) The maximum neighborhood size of $G_n$, $q_n = \max_{i=1,\ldots,p_n+1} |\mathrm{adj}_i(G_n)|$, satisfies $q_n = O(n^{1-b})$ for some $0 < b \leq 1$.
(E) The partial correlations $\rho_{nij|S}$ between $X_{ni}$ and $X_{nj}$ given $S$ satisfy the following upper and lower bounds, uniformly over $i, j \in \{1, \ldots, p_n + 1\}$ and $S \subseteq \{X_{n1}, \ldots, X_{n,p_n+1}\} \setminus \{X_{ni}, X_{nj}\}$:

$$\sup_{n, i \neq j, S} |\rho_{nij|S}| \leq M \qquad \text{for some } M < 1, \tag{6}$$

$$\inf_{i,j,S} \{|\rho_{nij|S}| : \rho_{nij|S} \neq 0\} \geq c_n, \tag{7}$$

where $c_n^{-1} = O(n^d)$ for some $0 < d < b/2$ with $b$ as in (D).
(F) The conditional variances satisfy the following bound:

$$\inf_{i=1,\ldots,p_n, S \subseteq \mathrm{adj}_i(G_n)} \frac{\mathrm{Var}(X_{ni}|S)}{\mathrm{Var}(Y_n|X_{ni}, S)} \geq v^2 \qquad \text{for some } v > 0.$$

Assumptions (C)–(E) were also made in [14]. Assumption (C) allows the number of covariates to grow as any polynomial of the sample size, representing the high-dimensional setting. Assumption (D) is a sparseness assumption requiring that the maximum neighborhood size in the DAG grows at a slower rate than $O(n)$. Condition (6) in assumption (E) excludes (sequences of) models in which the partial correlations approach 1, hence avoiding identifiability problems. Condition (7) in assumption (E) requires the nonzero partial correlations to be outside of the $n^{-b/2}$ range, with $b$ as in assumption (D). Note that this condition is similar to Assumption 5 in [21] and condition (8) in [36]. Finally, we note that assumption (F) is of the same spirit as Assumption 2 in [21]. Namely, if we scale $Y_n$ such that $\mathrm{Var}(Y_n) = \sigma^2$ for all $n$, then assumption (F) is implied by requiring that $\mathrm{Var}(X_{ni}|S) \geq v^2\sigma^2$ for all $i = 1, \ldots, p_n$ and $S \subseteq \mathrm{adj}_i(G_n)$.

Under assumptions (A) and (C)–(E), the PC-algorithm was shown to be consistent ([14], Theorem 2). The underlying reason for this result is the hierarchical nature of estimation and testing of partial correlations within the PC-algorithm. Due to sparsity and the faithfulness assumption, there is no need to estimate high-order partial correlations. This, together with the fact that the error in the estimation of partial correlations decays exponentially

16   M. H. MAATHUIS, M. KALISCH AND P. BÜHLMANNfast with increasing sample size, form the key elements of the consistency proof for the underlying CPDAG.

Consistency of the PC-algorithm means that there is a sequence $\alpha_n$ such that $P(\hat{G}_n(\alpha_n) = G_n) \to 1$ as $n \to \infty$. By combining this with the fact that for any given valid CPDAG the sample versions of Algorithms 1 and 3 perform exactly the same linear regressions, the following result is immediate.

THEOREM 5.1. *Under assumptions* (A) *and* (C)–(E), *there is a sequence* $\alpha_n$ *such that for all* $n \geq 1$ *the following holds on sets* $A_n$ *with* $P(A_n) \to 1$:

$$\hat{\Theta}_{ni}(\alpha_n) \stackrel{\text{set}}{=} \hat{\Theta}^L_{ni}(\alpha_n) \qquad \text{for all } i = 1, \ldots, p.$$

The next theorem shows that $\hat{\Theta}_{ni}$ and $\hat{\Theta}^L_{ni}$ are consistent estimators for $\Theta_{ni}$ and $\Theta^L_{ni}$, respectively.

THEOREM 5.2. *Under assumptions* (A) *and* (C)–(F), *there exists a sequence* $\alpha_n$ *such that*

$$\sup_{i=1,\ldots,p_n} d_{\text{multiset}}(\hat{\Theta}_{ni}(\alpha_n), \Theta_{ni}) \to_p 0,$$

$$\sup_{i=1,\ldots,p_n} d_{\text{multiset}}(\hat{\Theta}^L_{ni}(\alpha_n), \Theta^L_{ni}) \to_p 0,$$

*where, for any two multisets* $A = \{a_1, \ldots, a_m\}$ *and* $B = \{b_1, \ldots, b_q\}$ *with order statistics* $a_{(1)} \leq \cdots \leq a_{(m)}$ *and* $b_{(1)} \leq \cdots \leq b_{(q)}$,

$$d_{\text{multiset}}(A, B) = \begin{cases} \sup_{j=1,\ldots,m} |a_{(j)} - b_{(j)}|, & \text{if } m = q, \\ \infty, & \text{if } m \neq q. \end{cases}$$

The proof of Theorem 5.2 is given in Section 8. The key elements of the proof are similar to the ones in the consistency proof of the PC-algorithm. We only need to perform a limited number of low-order regression problems, and the estimation error we make in such problems decays exponentially fast when the sample size increases.

Figure 3 illustrates the connections between Theorems 3.2, 5.1 and 5.2. In particular, combining Theorems 3.2 and 5.2 yields that the elements of $\hat{\Theta}^L_{ni}$ converge in probability to elements of $\Theta_{ni}$, uniformly over the elements in $\hat{\Theta}^L_{ni}$ and $i = 1, \ldots, p_n$. Moreover, every element of $\Theta_{ni}$ is reached in this way. This leads to the following corollary.

COROLLARY 5.3. *Let* $f : \mathbb{R} \to \mathbb{R}$ *be a continuous function. Then, under assumptions* (A) *and* (C)–(F),

$$\sup_{i=1,\ldots,p_n} |\min\{f(\hat{\theta}) : \hat{\theta} \in \hat{\Theta}^L_{ni}\} - \min\{f(\theta) : \theta \in \Theta_{ni}\}| \to_p 0.$$

$$\begin{array}{ccc} \Theta^L_{ni} & \stackrel{\text{set}}{=} & \Theta_{ni} \\ \uparrow & & \uparrow \quad \text{(as multisets)} \\ \hat{\Theta}^L_{ni} & \stackrel{\text{set}}{=} & \hat{\Theta}_{ni} \quad \text{(on } A_n\text{)} \end{array}$$

FIG. 3. *Illustration of the connections between $\Theta^L_{ni}$, $\Theta_{ni}$, $\hat{\Theta}^L_{ni}$ and $\hat{\Theta}_{ni}$, given by Theorems 3.2, 5.1 and 5.2.*

An important implication of this corollary is obtained by taking $f(x) = |x|$, yielding that, under assumptions (A) and (C)–(F),

(8) $$\sup_{i=1,\ldots,p_n} |\min\{|\hat{\theta}|: \hat{\theta} \in \hat{\Theta}^L_{ni}\} - \min\{|\theta|: \theta \in \Theta_{ni}\}| \to_p 0.$$

The minimum absolute value of $\Theta_{ni}$ is a lower bound on the size of the causal effect of $X_i$ on $Y$. Equation (8) implies that we can estimate this bound consistently via the local method, uniformly in $i = 1, \ldots, p_n$.

Another implication of Corollary 5.3 follows by taking $f(x) = x$ and $f(x) = -x$, yielding that the local method is consistent for the joint estimation of $(\min(\Theta_{ni}), \max(\Theta_{ni})) = (\min\{\theta : \theta \in \Theta_{ni}\}, \max\{\theta : \theta \in \Theta_{ni}\})$, uniformly in $i = 1, \ldots, p_n$. Hence, any continuous function $g : \mathbb{R}^2 \to \mathbb{R}$ of $(\min(\Theta_{ni}), \max(\Theta_{ni}))$ can be consistently estimated by the local method. In particular, taking $g(x, y) = y - x$, we obtain that under assumptions (A) and (C)–(F)

$$\sup_{i=1,\ldots,p_n} |\text{range}(\hat{\Theta}^L_{ni}) - \text{range}(\Theta_{ni})| \to_p 0.$$

Thus, the range of possible causal effects of $X_i$ on $Y$ can be consistently estimated by the local method, uniformly in $i = 1, \ldots, p_n$.

We close this section by pointing out that not all functions of $\Theta_{ni}$ can be consistently estimated by the local method. For example, the mean of $\hat{\Theta}^L_{ni}$ is typically not a consistent estimate of the mean of $\Theta_{ni}$, since the multiplicities of $\Theta_{ni}$ and $\Theta^L_{ni}$ have different meanings (see Remark 3.3). In our simulations, however, the local method still yielded surprisingly good results in such a setting (see Figure 4, left panel).

**6. Simulations and real data analysis.** We now demonstrate the behavior of our methods in simulation studies and on a real data set. First, in Section 6.1, we use simulation studies to examine the behavior and speed of the basic method (Algorithm 1) and the local method (Algorithm 3). Next, in Section 6.2, we apply our methods to the problem of riboflavin production by *B. subtilis* that was discussed in the Introduction.

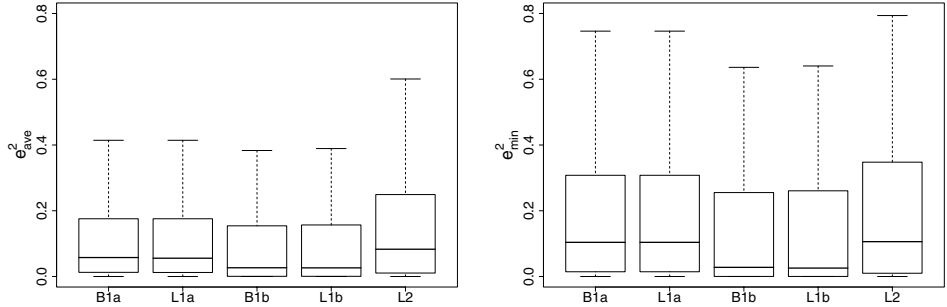

FIG. 4. *Comparison of the basic method* (B) *and the local method* (L) *over settings* 1(a), 1(b) *and* 2. *The left panel shows boxplots for* $e_{\text{ave}}^{2(k)}$ *and the right panel shows boxplots for* $e_{\text{min}}^{2(k)}$, $k = 1, \ldots, n_{\text{reps}}$ *(outliers excluded). The combination of the algorithm* (B/L) *and the simulation setting [1*(a)/1(b)/2*] is indicated on the x-axis.*

6.1. *Simulation studies.* We use the following simulation scheme. We generate $n_{\text{reps}}$ i.i.d. DAGs with edge weights for the following two settings:

Setting 1: $p + 1 = 10$, $\quad en = 4$, $\quad n_{\text{reps}} = 1000$,

Setting 2: $p + 1 = 1000$, $\quad en = 4$ (block structure), $\quad n_{\text{reps}} = 100$,

where $p + 1$ is the number of vertices of the DAG and $en$ is the expected neighborhood size of the DAG. The simulation of a single DAG with edge weights proceeds as follows. First, we use the R-package pcalg [15] to simulate a random DAG on $X_1, \ldots, X_{p+1}$ with the pre-specified expected neighborhood size $en$. In setting 2, we enforce a special block structure on the DAG by letting it consist of 100 disconnected components (blocks) of 10 variables each. Subsequently, we equip all edges $X_i \leftarrow X_j$ with edge weights $\beta_{ij}$ which are drawn independently from a Uniform([1, 2]) distribution.

For each $k = 1, \ldots, n_{\text{reps}}$ in the two settings, the DAG $G^{(k)}$ with edge weights $\beta_{ij}^{(k)}$ defines an underlying distribution on $(X_1^{(k)}, \ldots, X_{p+1}^{(k)})$:

(9)
$$\text{let } \varepsilon_1, \ldots, \varepsilon_{p+1}, \quad \text{i.i.d.} \sim \mathcal{N}(0, 1),$$
$$\text{for } i = 1, \ldots, p+1, \quad \text{set } X_i^{(k)} = \sum_{X_j^{(k)} \in \text{pa}_i(G^{(k)})} \beta_{ij}^{(k)} X_j^{(k)} + \varepsilon_i.$$

[Note that the $X_i^{(k)}$'s can be defined recursively as in (9), since pcalg automatically orders the variables in the DAGs so that $\text{pa}(X_1) = \varnothing$ and $\text{pa}_i \subseteq \{X_1, \ldots, X_{i-1}\}$ for $i = 2, \ldots, p+1$.]

For each DAG $G^{(k)}$, we randomly choose one vertex as the response variable $Y^{(k)}$ and another vertex as the covariate of interest $X^{(k)}$. We then determine the true multiset of possible causal effects of $X^{(k)}$ on $Y^{(k)}$ based on the true underlying distribution of $(X_1^{(k)}, \ldots, X_{p+1}^{(k)})$, and denote this by $\Theta^{(k)}$. In setting 2, $X^{(k)}$ and $Y^{(k)}$ are randomly chosen from the same block, in order to allow for a more direct and fair comparison with setting 1. [If $X^{(k)}$ and $Y^{(k)}$ were chosen from different blocks, then the causal effect could be quite easily identified as zero, giving an unfair advantage to setting 2.]

For each DAG $G^{(k)}$, we simulate a data set consisting of $n$ i.i.d. copies of $(X_1^{(k)}, \ldots, X_{p+1}^{(k)})$. We use two different sample sizes for setting 1, and one sample size for setting 2:

Setting 1: $n = 20$ [setting 1(a)] and $n = 2000$ [setting 1(b)],

Setting 2: $n = 100$.

Based on these simulated data, we compute estimates of $\Theta^{(k)}$, using tuning parameter $\alpha = 0.01$ in the PC-algorithm. In settings 1(a) and 1(b), we use both the basic and the local algorithm. In setting 2, we only use the local algorithm, since the basic algorithm is infeasible. We denote the output of the basic algorithm by $\hat{\Theta}^{(k)}$ and the output of the local algorithm by $\hat{\Theta}^{(k,L)}$.

We compare $\hat{\Theta}^{(k)}$ to $\Theta^{(k)}$ using the following two measures:

$$e_{\text{ave}}^{2(k)} = \left(|\hat{\Theta}^{(k)}|^{-1} \sum_{\hat{\theta} \in \hat{\Theta}^{(k)}} |\hat{\theta}| - |\Theta^{(k)}|^{-1} \sum_{\theta \in \Theta^{(k)}} |\theta|\right)^2,$$

$$e_{\text{min}}^{2(k)} = (\min\{|\hat{\theta}| : \hat{\theta} \in \hat{\Theta}^{(k)}\} - \min\{|\theta| : \theta \in \Theta^{(k)}\})^2$$

with analogous measures for comparing $\hat{\Theta}^{(k,L)}$ to $\Theta^{(k)}$. Note that $e_{\text{ave}}^{2(k)}$ measures the squared error in the estimation of the mean absolute value of $\Theta^{(k)}$, and $e_{\text{min}}^{2(k)}$ measures the squared error in the estimation of the minimum absolute value of $\Theta^{(k)}$.

Figure 4 compares the results of the basic method and the local method, showing boxplots for $e_{\text{ave}}^2$ (left panel) and $e_{\text{min}}^2$ (right panel). From the discussion following Corollary 5.3, we know that the local method is consistent for the minimum absolute value of $\Theta^{(k)}$, while it is typically inconsistent for the mean absolute value of $\Theta^{(k)}$. On the other hand, the basic method is consistent for both parameters. In light of this, it is surprising to see that the boxplots for the basic method and the local method are basically identical for both measures of performance $e_{\text{ave}}^2$ and $e_{\text{min}}^2$. We also note that both methods perform better in setting 1(b) than in setting 1(a) because of the larger sample size in setting 1(b). Finally, the performance of the local method deteriorates only slightly in the high-dimensional setting 2.

In order to demonstrate the behavior of the basic method and the local method in more detail, we also evaluate their performance on several data sets that are generated from a fixed DAG with edge weights. Thus, we generate a random DAG $G$ ($p = 7, en = 3$) with edge weights, and randomly choose a covariate $X$ and a response variable $Y$, as before. Next, we generate 50 data sets of size 1000 from this DAG, according to the model given in (9). For each data set, we estimate the multiset of possible causal effects, using $\alpha = 0.01$. We then aggregate these 50 estimates, and construct a density plot.

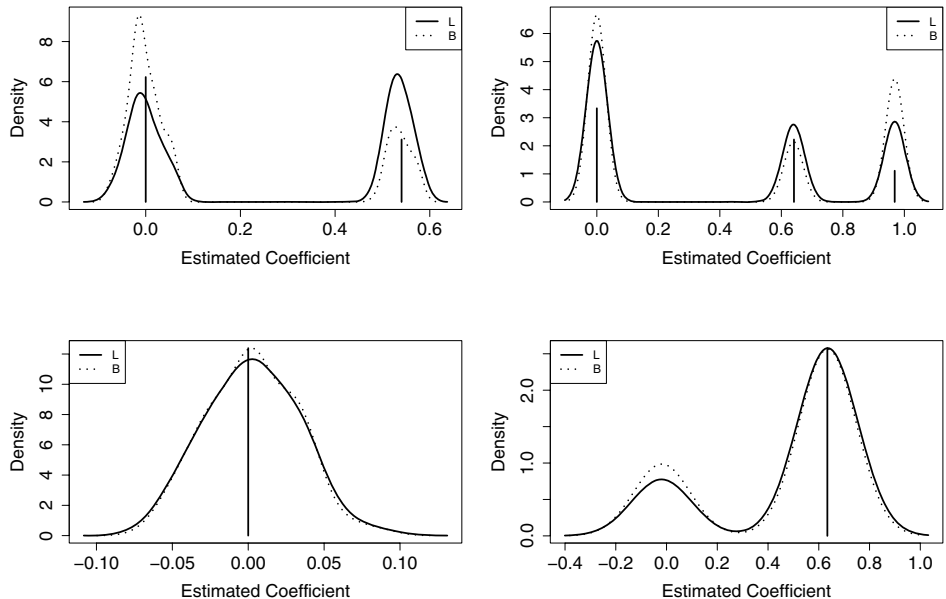

FIG. 5. *The estimated effects (density plots for the output of the basic and the local method over 50 replicates) are compared to the true multisets of possible causal effects (vertical lines; heights indicate the relative frequencies of the values). The parameters in all four settings are $p = 7$, $en = 3$, $n = 1000$, $\alpha = 0.01$.*

Figure 5 shows the results for four typical DAGs. The true multisets of possible causal effects are indicated by vertical lines, where the height of each line indicates the relative frequency of the given value in the multiset. In the upper left panel, we see that both methods pick up the set of possible causal effects quite reliably. The basic method captures the multiplicities better than the local method, as expected from our theory (see Remark 3.3). However, this advantage of the basic method is not so clear in the upper right panel. The lower left panel shows an example where the true causal effect is zero, and this is identified correctly by both methods. Finally, the lower right panel shows an example where the true causal effect is unique and is approximately 0.63. Both methods find this effect, but they also identify zero as a possible causal effect. This error is caused by the fact that the CPDAG is estimated incorrectly for some of the 50 data sets.

Finally, we consider the runtime of the algorithms. Table 1 shows that the runtime of the basic algorithm is much larger and much more volatile than the runtime of the local algorithm. This was to be expected, since the basic algorithm has to find all DAGs within an equivalence class. In our implementation, graphs with 15 vertices or more cannot be handled reliably by the basic algorithm, while they can be handled easily by the local algorithm.

TABLE 1
*Mean runtime in seconds of the basic algorithm and the local algorithm over 10 replicates with settings $en = 3$, $n = 1000$, $\alpha = 0.01$, and the specified number of covariates p. Standard errors of the mean are given in parentheses. A value NA means that at least one of the 10 replicates took more than 48 hours to compute, so that the computation was aborted. All computations were carried out on a 2.6 GHz Dual-Core AMD Opteron processor with 32 GB RAM on red hat Linux 2.6.18, using R 2.7.2*

|       | $p = 4$       | $p = 9$       | $p = 14$    | $p = 29$    | $p = 49$    | $p = 99$ |
|-------|---------------|---------------|-------------|-------------|-------------|----------|
| Basic | 0.120 (0.01)  | 17.6 (5.4)    | NA          | NA          | NA          | NA       |
| Local | 0.038 (0.002) | 0.088 (0.008) | 0.15 (0.02) | 0.50 (0.06) | 0.99 (0.06) | 2.8 (0.3) |

6.2. *Riboflavin data.* We now apply our methods to a data set about riboflavin (vitamin $B_2$) production by *B. subtilis*, kindly provided to us by DSM Nutritional Products (Switzerland). As discussed in the Introduction, the data are observational. The real-valued response variable is the logarithm of the riboflavin production rate, and there are $p = 4088$ covariates measuring the logarithm of the expression level of 4088 genes that cover essentially the whole genome of *Bacillus subtilis*. The sample size is $n = 71$, and, hence, this is a high-dimensional setting with $p \gg n$.

The data are of high quality, for example, in terms of a large signal to noise ratio in a properly regularized linear model. Furthermore, Gaussianity of the marginal distributions of the data seems a reasonable approximation. Detecting strong deviations from joint multivariate Gaussianity in such high-dimensional data is extremely hard, as is verification of the DAG and faithfulness assumptions. A more detailed discussion about these assumptions can be found in Section 7.

Due to the large number of covariates in this data set, our basic algorithm is infeasible, and we only apply the local algorithm. After standardizing the data so that all covariates have unit variance, we estimate the multiset of possible causal effects of each gene on the riboflavin production. We first analyze the number of distinct values in each of these multisets, which we call the *ambiguity* of the multiset. In high-dimensional problems, one might fear that these ambiguities can be very large, but this is not the case for the riboflavin data. Varying the tuning parameter $\alpha$ for the PC-algorithm between 0.01 and 0.5, there is no gene in the pool of 4088 genes with an ambiguity greater than 5, and the large majority of genes have ambiguity 1 (i.e., they yield a unique estimate for the causal effect; see Table 2).

Table 2
*The fraction of the 4088 genes in the riboflavin data set with a certain ambiguity â, for various values of the tuning parameter $\alpha$*

|  | $\hat{a} = 1$ | $\hat{a} = 2$ | $\hat{a} = 3$ | $\hat{a} = 4$ | $\hat{a} = 5$ |
| --- | --- | --- | --- | --- | --- |
| $\alpha = 0.01$ | 0.775 | 0.186 | 0.036 | 0.004 | 0.001 |
| $\alpha = 0.05$ | 0.845 | 0.120 | 0.029 | 0.005 | 0.001 |
| $\alpha = 0.1$ | 0.897 | 0.085 | 0.016 | 0.002 | 0 |
| $\alpha = 0.2$ | 0.951 | 0.042 | 0.005 | 0.002 | 0 |
| $\alpha = 0.3$ | 0.970 | 0.025 | 0.003 | 0.002 | 0 |
| $\alpha = 0.4$ | 0.974 | 0.023 | 0.002 | 0.001 | 0 |
| $\alpha = 0.5$ | 0.981 | 0.018 | 0.001 | 0 | 0 |

In the remainder of the analysis, we set the tuning parameter $\alpha$ to 0.01. In order to obtain a single estimate for the causal effect of each gene, we compute the minimum absolute value of its estimated multiset. As discussed before, this is a consistent estimate for the minimum absolute value of the true multiset of possible causal effects of the gene (under our assumptions). In order to assess the reliability of these estimates, we bootstrap the data 10 times and take the median of the 10 estimates for each gene. We call the resulting values the *causal scores* of the genes. Figure 6 shows a histogram of these causal scores. Note that the histogram has a strong right tail, indicat-ing that there is a group of genes with strongly estimated causal effects that are stable in a bootstrap analysis. In order to decide which causal scores should be considered "significantly high," we use the local false discovery rate (FDR) [7]. The vertical line in Figure 6 shows the cut-off for a local FDR of 10%. About 200 of the 4088 genes fall to the right of this cut-off, and hence have a local FDR that is less than 10%. According to our analysis, these genes are promising candidates for genetic modification.

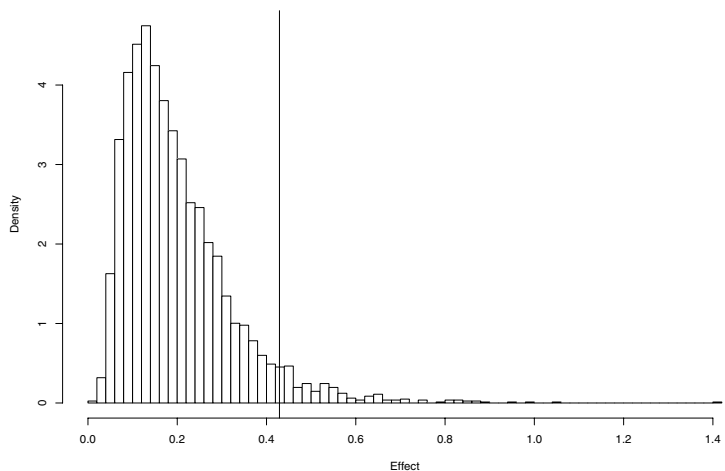

Fig. 6. *Histogram of the causal scores (median of the minimum absolute effect over 10 bootstrap samples) for the 4088 genes in the riboflavin data set. All genes to the right of the vertical line have a local FDR of less than* 10%.

We compare our method to an association approach using regression, which is, as we have argued before, inappropriate for inferring causal effects. To cope with high-dimensional variable selection in a linear model, we use the (prediction optimal tuned) Lasso; properties of the Lasso for variable selection in regression are discussed in [21] and [36]. Among the top ten genes of Lasso (ordered by absolute values of estimated regression coefficients), we found only one gene that was also among the top ten genes of our method (ordered by the causal scores). This difference is due to the fact that causal effects and associations can be very different. To evaluate whether the results of our intervention approach are superior in practice, we would need to compare the intervention effects that we computed from the observational data to intervention effects computed from lab experiments in which the interventions were actually carried out. We are currently in the process of doing this. However, already at this point we can say that if the target is prediction of intervention effects or causal effects, an association analysis like regression will not provide an appropriate answer.

**7. Discussion.** In this paper, we present a new method that combines estimation of the equivalence class of DAGs with causal inference methods that can be used when the DAG is known. The need for such a combination is due to the fact that, for a large class of practical problems, it is unrealistic to assume that the graph structure among the variables of interest is known. Thus, we assume that we have observational data that were generated from an *unknown* DAG, and based on these data we want to estimate causal effects. We argue that, in this situation, causal effects can typically not be uniquely determined, and we focus our estimation on the multisets $\Theta_i$ of possible causal effects of $X_i$ on $Y$, $i = 1, \ldots, p$. Summary measures of $\Theta_i$ can be used to determine variable importance. In particular, we propose using the minimum absolute value of $\Theta_i$, since this is a lower bound on the size of the causal effect of $X_i$ on $Y$. We show that the distinct values of $\Theta_i$ can be estimated by a fast *local* method, which is computationally feasible and asymptotically consistent in sparse high-dimensional settings. Thus, we achieve consistent estimation, based on observational data, for the lower bound of the size of each individual covariate's total causal effect on $Y$.

The motivation for our work comes from a problem about genetic engineering of *Bacillus subtilis* in order to improve its riboflavin production rate. The response variable of interest is the riboflavin production rate, and there are $p = 4088$ covariates (genes) from which we have expression levels. Based on these observational data, our goal is to find genes that are good candidates for single-gene interventions that improve the riboflavin production rate. With our new method, we find a list of genes whose top-ranking members are surprisingly stable (when doing a bootstrap analysis) and clearly relevant in terms of a local false discovery rate. Furthermore, our list of genes with large lower bounds for their causal effects is markedly different from a regression approach, which measures only association (instead of intervention or causality).

One should be careful not to over-interpreting our results. We have shown that, within the class of Gaussian distributions that are faithful to the true (unknown) causal DAG, it is possible to estimate good lower bounds for causal effects on the basis of observational data. In practice, it is hard or impossible to check whether our assumptions hold, at least in an approximate sense.

The Gaussian assumption is conceptually not a key assumption. For non-Gaussian data, the PC-algorithm can still be used to estimate the equivalence class of DAGs, and the causal effects are still given by (3) [but they will not be constant in general, and depend on the value of $x'_i$ in (3)]. It is also interesting that in a certain sense the Gaussian assumption makes things more difficult. If the linearity assumption is retained and the errors are assumed to be *non*-Gaussian, then the DAG can be *uniquely* recovered [29], preventing the need to work with equivalence classes. We note, however, that Gaussianity is essential for our consistency proofs of the algorithms in high-dimensional settings. Consistency of the PC-algorithm for high-dimensional non-Gaussian data is still an open problem, and consistent estimation of the causal effects also seems nontrivial in general, involving function estimation.

The DAG assumption implicitly includes the assumption that we have no unmeasured confounders. This is a very strong assumption in general. In our particular example from biology, this assumption may be reasonable though, since we observe the expression levels from essentially all genes of the *Bacillus subtilis* genome (but of course there may still be some unobserved aspects of the genome). The presence of hidden variables can lead to various problems. First, since the class of graphical Markov models is not closed under marginalization [26], it can happen that there is no CPDAG that represents all and only all conditional independence relationships among the observed variables. This causes problems in the PC-algorithm with conflicting $v$-structures, even for infinite sample sizes. Hence, the presence of conflicting $v$-structures for data sets where $n$ is large relative to $p$ may be interpreted as a warning sign for the presence of hidden variables. But even without conflicting $v$-structures, hidden variables can lead to erroneous conclusions. To see this, suppose that data are generated according to the DAG in Figure 7(a), and that variable $L_1$ is not observed. Then, the causal effect of $X_1$ on $Y$ is not identifiable, because any part of the observed correlation between $X_1$ and $Y$ can be explained by $L_1$. However, applying the PC-algorithm to the distribution of the observed variables $(X_1, Y)$ yields the CPDAG $X_1 - Y$, and our approach wrongly concludes that the causal effect of $X_1$ on $Y$ is contained in the set $\{0, \beta_{1|\varnothing}\}$. As a second example, suppose that data are generated according to the DAG in Figure 7(b) and that variables $L_1$ and $L_2$ are not observed. From the DAG, it is clear that neither $X_1$ nor $X_2$ have a causal effect on $Y$. But the distribution of the observed variables $(X_1, X_2, Y)$ is faithful to the DAG $X_1 \to Y \leftarrow X_2$. Hence, applying the PC-algorithm to the distribution of the observed variables yields this DAG, and our approach wrongly concludes that the causal effects of $X_1$ on $Y$ and $X_2$ on $Y$ are unique and equal to $\beta_{1|\varnothing}$ and $\beta_{2|\varnothing}$, respectively. The problem in these two examples is that, although the CPDAGs $X_1 - Y$ and $X_1 \to Y \leftarrow X_2$ represent all and only all conditional independence relationships among the observed variables, these CPDAGs may no longer be interpreted causally. Relaxing the assumption of unmeasured confounders is possible by extending our methodology to ancestral graphs (see [26, 27] and [34]) which allow for hidden variables. However, deriving bounds for causal effects when the underlying ancestral graph is unknown is an open issue.

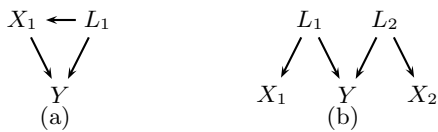

FIG. 7. *Examples illustrating that our approach can lead to erroneous conclusions in the presence of hidden variables.*

Another interesting direction of future research consists of considering other types of causal effects than the total effect of a single covariate on a response, such as the direct effect of a single covariate on a response or the total effect of a joint intervention on several covariates. The conceptual idea of our basic algorithm (Algorithm 1) can be used for any type of causal effect. But our local algorithm (Algorithm 3) relies on the fact that the total effect of a covariate $X$ on a response $Y$ can be computed if one knows the parents of $X$. Such a simple relationship does not hold in general for other types of causal effects, and it would be interesting to investigate for which other types of causal effects a local approach like the one in Algorithm 3 could be used.

We conclude by coming back to our practical problem of riboflavin production by *Bacillus subtilis*. This problem is of a causal or interventional type, and, hence, our intervention approach is more appropriate than a regression-type association analysis using high-dimensional variable selection in a linear model. Therefore, despite open issues in the difficult field of inferring bounds for causal effects, our new approach offers both conceptual and practical improvements.

**8. Proofs.** In order to prove Lemma 3.1, we need some more graph theory and terminology. Consider an undirected graph $G = (V, E)$. For any subset $V' \subseteq V$, the *subgraph induced by* $V'$ is $G_{V'} = (V', E_{V'})$, where $E_{V'}$ is the set of all edges in $E$ whose endpoints are both in $V'$. $G$ is called *chordal* (or *triangulated*) if each of its cycles of length four or more has a *chord*, which is an edge joining two nonconsecutive vertices in the cycle. $G$ is called *complete* if every pair of distinct vertices is adjacent. A *clique* is a set of vertices so that every pair of distinct vertices in this set is adjacent. A vertex is *simplicial* if its adjacency set forms a clique. A *perfect elimination scheme* of a graph $G$ is an ordering $\sigma = (v_1, \ldots, v_n)$ of its vertices, so that each $v_i$ is a simplicial vertex in the induced subgraph $G_{\{v_i,\ldots,v_n\}}$.

Chordal graphs have many nice properties. We will use the following (cf. [1, 6] and [9]):

1. Every chordal graph $G$ has a simplicial vertex. If $G$ is not complete, then it has at least two nonadjacent simplicial vertices.
2. Chordality of graphs is a hereditary property: if $G = (V, E)$ is chordal, then all subgraphs of $G$ induced by $V' \subseteq V$ are chordal.
3. Every chordal graph has a perfect elimination scheme.

We also note that we can turn an undirected graph into a DAG without $v$-structures by directing its edges according to a perfect elimination scheme $\sigma = (v_1, \ldots, v_n)$: for any vertex $v_i$, determine the adjacency set of $v_i$ in $G_{\{v_i,\ldots,v_n\}}$, and for each vertex $v_j$ in this adjacency set, direct the edge $v_j$–$v_i$ toward $v_i$. Note that the ordering of the vertices ensures that we cannot create directed cycles. Moreover, we cannot create $v$-structures, since the adjacency set of $v_i$ in $G_{\{v_i,\ldots,v_n\}}$ is a clique for all $i = 1, \ldots, n$.

PROOF OF LEMMA 3.1. Let $i \in \{1, \ldots, p\}$, and let $S \subseteq \text{sib}_i(G)$. We only prove the nontrivial direction of the lemma. We assume that $G_{S \to i}$ is locally valid, and we show that there is a corresponding DAG $G^*$ in the equivalence class with $\text{pa}_i(G^*) = \text{pa}_i(G) \cup S$.

First, we note that $X_i \cup S$ is a clique. This is trivial if $S = \varnothing$. If $S \neq \varnothing$, take an arbitrary vertex $v$ in $S$. Since $S \subseteq \text{sib}_i(G)$, $v$ is adjacent to $X_i$. It must also be adjacent to all other vertices in $S$, since, otherwise, $G_{S \to i}$ contains a new $v$-structure with $X_i$ as collider, and this contradicts the assumption that $G_{S \to i}$ is locally valid.

Next, we use the following facts that were proved in [20], Proof of Theorem 3: (i) no orientation of the edges not oriented in $G$ will create a directed cycle which includes an edge or edges that were oriented in $G$, (ii) no orientation of an edge not directed in $G$ can create a new $v$-structure with an edge that was oriented in $G$ and (iii) the subgraph $G'$ of $G$, obtained by removing all of the oriented edges in $G$, is the union of disjoint chordal graphs. Combining these facts implies that any orientation of the edges in $G'$ that does not create directed cycles or $v$-structures corresponds to a DAG in the equivalence class of $G$. Moreover, in order to find such an orientation, each of the disjoint chordal graphs in $G'$ can be considered separately.

Let $G_1'', \ldots, G_d''$ be the collection of disjoint chordal graphs constituting $G'$. Without loss of generality, we assume that $X_i$ is contained in $G_1''$. Since $G_2'', \ldots, G_d''$ are chordal, we can find a perfect elimination scheme for each of these graphs and order their edges accordingly. We need to be more careful with $G_1''$, since we need to find a direction of the edges so that all and only all vertices in $S$ are parents of $X_i$. In terms of a perfect elimination scheme, this means that we need to order the vertices $V_1''$ of $G_1''$, such that all vertices in $V_1'' \setminus (\{X_i \cup S\})$ are ordered before $X_i$, and all vertices in $S$ are ordered after $X_i$. If $G_1''$ is complete, then such an ordering exists trivially, since any ordering of the vertices of a complete graph is a perfect elimination scheme. If $G_1''$ is not complete, then there must be at least two nonadjacent simplicial vertices. Since $\{X_i\} \cup S$ is a clique, at least one of these vertices must be in $V_1'' \setminus (\{X_i \cup S\})$. We take such a vertex, say $v_1$, as the first vertex in the perfect elimination scheme. Next, we consider the induced subgraph $G_{V_1'' \setminus \{v_1\}}$. This graph is again chordal, since chordality is a hereditary property. If $G_{V_1'' \setminus \{v_1\}}$ is complete, then we are done. If it is not complete, then we take a simplicial vertex in $(V_1'' \setminus \{v_1\}) \setminus (\{X_i\} \cup S)$ as the next vertex in the elimination scheme. We repeat this procedure until it terminates, which is guaranteed to happen for some graph $G_A$ with $A \supseteq (\{X_i\} \cup S)$, since $\{X_i\} \cup S$ is a clique.

Directing the edges of $G_1'', \ldots, G_d''$ according to the resulting perfect elimination schemes yields a DAG without $v$-structures and with the same skeleton as $G'$, where all and only all vertices in $S$ are parents of $X_i$. Hence, using this direction of edges in the original CPDAG $G$ yields a DAG $G^*$ that is in the equivalence class of $G$ and that satisfies $\mathrm{pa}_i(G^*) = \mathrm{pa}_i(G) \cup S$. □

In order to prove Theorem 5.2, we need the following lemma.

LEMMA 8.1. *Assume that assumptions* (A) *and* (C)–(F) *hold. Then, for every* $\varepsilon > 0$, *we have*

$$\sup_{i=1,\ldots,p_n, S \subseteq \mathrm{adj}_i(G_n)} P(|\hat{\beta}_{ni|S} - \beta_{ni|S}| > \varepsilon)$$
$$\leq \frac{C_1}{\varepsilon} \exp(-C_2 \varepsilon^2 (n - q_n - 1)) + 2\exp(-C_3(n/2 - q_n - 1)), \qquad n \geq N,$$

where $N>0$ is a constant depending on $q_n$ [see assumption (D)], $C_1, C_2 > 0$ are constants depending on $v$ [see assumption (F)], and $C_3 > 0$ is an absolute constant.

PROOF. Let $i \in \{1, \ldots, p_n\}$, $S \subseteq \mathrm{adj}_i(G_n)$, and $\varepsilon > 0$. If $Y_n \in S$, then $\hat{\beta}_{ni|S} = \beta_{ni|S} = 0$. Hence, we assume $Y_n \notin S$. In that case $\hat{\beta}_{ni|S}$ is the estimated regression coefficient of $X_{ni}$ in the regression of $Y_n$ on $X_{ni}$ and $S$, and $\beta_{ni|S}$ is the true regression coefficient of $X_{ni}$ in the regression of $Y_n$ on $X_{ni}$ and $S$.

We first analyze the distribution of $\hat{\beta}_{ni|S}|\{X_{ni}, S\}$. Let $\sigma^2_{ny|i,S}$ denote the variance of $Y_n|\{X_{ni}, S\}$, and let $\sigma^2_{ni|S}$ denote the variance of $X_{ni}|S$. Moreover, let $s^2_{ni}$ denote the sample variance of $X_{ni}$ [using $(n-1)$ in the denominator], let $s^2_{ni|S}$ denote the sample variance of $X_{ni}|S$ (using the residuals in the regression of $X_{ni}$ on $S$, with $n - |S| - 1$ in the denominator), and let $R^2_{ni|S}$ denote the sample multiple correlation coefficient between $X_{ni}$ and $S$. Then,

$$(10) \quad \mathrm{Var}(\hat{\beta}_{ni|S}|\{X_{ni}, S\}) = \frac{1}{1 - R^2_{ni|S}} \frac{\sigma^2_{ny|i,S}}{(n-1)s^2_{ni}} = \frac{\sigma^2_{ny|i,S}}{(n-|S|-1)s^2_{ni|S}},$$

where the first equality is a well-known identity, and the second equality follows from $1 - R^2_{ni|S} = \{(n - |S| - 1)s^2_{ni|S}\}/\{(n-1)s^2_{ni}\}$. Combining (10) with $E(\hat{\beta}_{ni|S}|\{X_{ni}, S\}) = \beta_{ni|S}$ and assumption (A), we obtain

$$(11) \quad P(|\hat{\beta}_{ni|S} - \beta_{ni|S}| > \varepsilon | \{X_{ni}, S\}) = P\left(|Z| > \frac{\varepsilon \sqrt{n - |S| - 1}\, s_{ni|S}}{\sigma_{ny|i,S}} \Big| \{X_{ni}, S\}\right),$$

where $Z$ is a standard normal random variable.

We first analyze (11) on the event $B_{niS} = \{X_{ni}, S : s^2_{ni|S} > \frac{1}{2}\sigma^2_{ni|S}\}$. Using assumption (F), we obtain

$$P\left(|Z| > \frac{\varepsilon \sqrt{n - |S| - 1}\, s_{ni|S}}{\sigma_{ny|i,S}} \Big| \{X_{ni}, S\}\right) 1_{B_{niS}}$$

$$\leq P\left(|Z| > \varepsilon v \frac{\sqrt{n - |S| - 1}}{\sqrt{2}}\right)$$

$$\leq P(|Z| > C\varepsilon \sqrt{n - q_n - 1}),$$

where $C$ depends on $v$, and $q_n$ is as in assumption (D). Using the well-known bound on tail probabilities of the standard normal distribution $P(|Z| > a) \leq 2/(\sqrt{2\pi}a) \exp(-a^2/2)$, the last display is bounded above by

$$\frac{C_1}{\varepsilon} \exp(-C_2 \varepsilon^2 (n - q_n - 1))$$

for all $n \geq q_n + 2$, where $C_1, C_2 > 0$ are constants depending on $v$.

Next, we compute an upper bound for $P(B_{niS}^C)$. Note that

$$P(B_{niS}^C | S) = P\left(\frac{(n - |S| - 1)s_{ni|S}^2}{\sigma_{ni|S}^2} \leq \frac{1}{2}(n - |S| - 1)|S|\right)$$
$$= P(\chi_{n-|S|-1}^2 \leq (n - |S| - 1)/2 | S)$$
$$\leq P(\chi_{n-q_n-1}^2 \leq (n - 1)/2),$$

where $\chi_k^2$ denotes a chi-squared random variable with $k$ degrees of freedom. We now apply Bernstein's inequality ([32], Lemma 2.2.11, page 103) by writing

$$P(\chi_{n-q_n-1}^2 \leq (n-1)/2) = P(\chi_{n-q_n-1}^2 - (n - q_n - 1) \leq -(n-1)/2 + q_n)$$
$$\leq P(|\chi_{n-q_n-1}^2 - (n - q_n - 1)| \geq (n-1)/2 - q_n).$$

By viewing a $\chi_{n-q_n-1}^2 - (n - q_n - 1)$ random variable as the sum of $n - q_n - 1$ independent centered $\chi_1^2$ random variables and noting that a centered $\chi_1^2$ random variable satisfies the moment conditions of Bernstein's inequality, it follows that the last display is bounded above by

$$2\exp\left(-\frac{((n-1)/2 - q_n)^2}{C_3' + C_4'((n-1)/2 - q_n)}\right),$$

where $C_3', C_4' > 0$ are constants coming from the moment conditions. This expression is in turn bounded above by $2\exp(-C_3(n/2 - q_n - 1))$ for all $n$ such that $(n-1)/2 - q_n \geq C_3'$. Since this bound holds for all $S$ with $|S| \leq q_n$, it also holds for the unconditional probability $P(B_{niS}^C)$.

The result now follows from putting the parts together:

$$P(|\hat{\beta}_{ni|S} - \beta_{ni|S}| > \varepsilon)$$
$$\leq \int P(|\hat{\beta}_{ni|S} - \beta_{ni|S}| > \varepsilon | \{X_{ni}, S\}) 1_{B_{niS}} dF_{X_{ni},S} + P(B_{niS}^C)$$
$$\leq \frac{C_1}{\varepsilon} \exp(-C_2 \varepsilon^2 (n - q_n - 1)) + 2\exp(-C_3(n/2 - q_n - 1)),$$

where $F_{X_{ni},S}$ denotes the distribution of $(X_{ni}, S)$. □

PROOF OF THEOREM 5.2. Let $\varepsilon > 0$. By consistency of the PC-algorithm ([14], Theorem 2), there is a sequence $\alpha_n$, such that $P(A_n) \to 1$ for the event $A_n = \{\hat{G}_n(\alpha_n) = G_n\}$. Hence, it is sufficient to show that

(12) $$\lim_{n \to \infty} P\left(\sup_{i=1,\ldots,p_n} d_{\text{multiset}}(\hat{\Theta}_{ni}(\alpha_n), \Theta_{ni}) > \varepsilon, A_n\right) \to 0 \quad \text{and}$$

(13) $$\lim_{n \to \infty} P\left(\sup_{i=1,\ldots,p_n} d_{\text{multiset}}(\hat{\Theta}_{ni}^L(\alpha_n), \Theta_{ni}^L) > \varepsilon, A_n\right) \to 0.$$

In the remainder of the proof, we suppress the dependence of $\alpha_n$ in the notation. We first consider the local method. On the event $A_n$, the cardinalities $|\hat{\Theta}_{ni}^L|$ and $|\Theta_{ni}^L|$ of the multisets $\hat{\Theta}_{ni}^L$ and $\Theta_{ni}^L$ are equal. Hence,

$$
(14) \quad \begin{aligned} & P\Big(\sup_{i=1,\ldots,p_n} d_{\text{multiset}}(\hat{\Theta}_{ni}^L, \Theta_{ni}^L) > \varepsilon, A_n\Big) \\ &= P\Big(\sup_{i=1,\ldots,p_n, j=1,\ldots,|\Theta_{ni}^L|} |\hat{\theta}_{ni(j)}^L - \theta_{ni(j)}^L| > \varepsilon, A_n\Big), \end{aligned}
$$

where $\hat{\theta}_{ni(j)}^L$ and $\theta_{ni(j)}^L$ are the order statistics of $\hat{\Theta}_{ni}^L$ and $\Theta_{ni}^L$, respectively. Moreover, on the event $A_n$, we have that for every $i=1,\ldots,p_n$ and $j=1,\ldots,|\Theta_{ni}^L|$, $\hat{\theta}_{ni(j)}^L = \hat{\beta}_{ni|\text{pa}_i(G_n)\cup S}$ for some $S \subseteq \text{sib}_i(G_n)$. Hence, $\theta_{ni(j)}^L = \hat{\beta}_{ni|S'}$ for some $S' \subseteq \text{adj}_i(G_n)$. Note, however, that $\hat{\theta}_{ni(j)}^L$ and $\theta_{ni(j)}^L$ do not need to correspond to the same set $S$, since it may happen that $\hat{\theta}_{ni(j)}^L = \hat{\beta}_{ni|S}$, $\theta_{ni(j)}^L = \beta_{ni|S'}$, and $\beta_{ni|S} \neq \beta_{ni|S'}$. But, since the pairing of the elements of $\hat{\Theta}_{ni}^L$ and $\Theta_{ni}^L$, with respect to their order statistics, is an optimal pairing for the supremum distance, the following inequality holds for all $i=1,\ldots,p_n$:

$$
\sup_{j=1,\ldots,|\Theta_{ni}^L|} |\hat{\theta}_{ni(j)}^L - \theta_{ni(j)}^L| \leq \sup_{S \subseteq \text{adj}_i(G_n)} |\hat{\beta}_{ni|S} - \beta_{ni|S}|.
$$

Combining this with (14) yields

$$
(15) \quad \begin{aligned} & P\Big(\sup_{i=1,\ldots,p_n} d_{\text{multiset}}(\hat{\Theta}_{ni}^L, \Theta_{ni}^L) > \varepsilon, A_n\Big) \\ &\leq P\Big(\sup_{i=1,\ldots,p_n, S \subseteq \text{adj}_i(G_n)} |\hat{\beta}_{ni|S} - \beta_{ni|S}| > \varepsilon\Big) \\ &\leq \sum_{i=1}^{p_n} \sum_{S \subseteq \text{adj}_i(G_n)} P(|\hat{\beta}_{ni|S} - \beta_{ni|S}| > \varepsilon) \\ &\leq p_n 2^{q_n} \sup_{i=1,\ldots,p_n, S \subseteq \text{adj}_i(G_n)} P(|\hat{\beta}_{ni|S} - \beta_{ni|S}| > \varepsilon), \end{aligned}
$$

where the last inequality follows from the fact that the number of possible subsets of $\text{adj}_i(G_n)$ is bounded above by $2^{q_n}$, where $q_n$ is given in assumption (D). Using Lemma 8.1 and assumptions (C) and (D), it follows that expression (15) converges to zero as $n \to \infty$. This completes the proof of (13), yielding consistency of the local method.

We can use the same reasoning for the basic method. To see this, we note that on the event $A_n$ the estimated CPDAG is a valid CPDAG. Hence, the sample versions of the basic and the local algorithm perform exactly the same linear regressions (cf. Theorem 5.1). The only difference in the output of the two algorithms lies in the multiplicities of the values. But since the estimated CPDAG is correct, the multiplicities of the sample version of the basic algorithm are correct, and they do not affect expression (12). □

## APPENDIX: POSSIBLE MODIFICATIONS OF THE ALGORITHMS

We first introduce some new notation. Let $\mathrm{pa}_{i,y}(G)$ be the set of vertices in $\mathrm{pa}_i(G)$ from which there is a path to $Y$. Similarly, let $\mathrm{sib}_{i,y}(G)$ be the set of vertices in $\mathrm{sib}_i(G)$ from which there is a path to $Y$.

We now discuss two modifications that can be made to the basic Algorithm 1:

(i) Replace line 4 of Algorithm 1 by: "If $G_j$ does not contain a directed path from $X_i$ to $Y$, then set $\theta_{ij} = 0$. Otherwise, set $\theta_{ij} = \beta_{i|\mathrm{pa}_i(G_j)}$." Since the causal effect of $X_i$ on $Y$ is by definition zero if there is no directed path from $X_i$ to $Y$, this modification does not change the output of the population version of the algorithm. In the sample version, however, it allows us to estimate exact zeroes, eliminating estimation error from the regression estimates when there is no directed path.

(ii) Replace $\mathrm{pa}_i(G_j)$ in line 4 of Algorithm 1 by $\mathrm{pa}_{i,y}(G_j)$. Since both $\mathrm{pa}_i(G_j)$ and $\mathrm{pa}_{i,y}(G_j)$ satisfy the back-door criterion with respect to $(X_i, Y)$, the output of the population version of the algorithm is again unchanged. In the sample version, this modification can be used to reduce the dimensionality of the regression problems.

One can also make several modifications to the local Algorithm 3:

(i) Before line 2 of Algorithm 3, test whether the CPDAG $G$ allows a directed path from $X_i$ to $Y$ (i.e., test whether it is possible to direct the undirected edges of $G$ so that a directed path from $X_i$ to $Y$ is created, without creating additional $v$-structures or directed cycles). If $G$ does not allow such a path, set $\Theta_i = \{0\}$. If $G$ does allow such a path, perform lines 2–7 of Algorithm 3. This modification may change the output of the population version of the algorithm, in the sense that the cardinality of $\Theta_i$ may change if $G$ does not allow a directed path. In such a case, the cardinality is always 1 in the modified version, while it equals the number of sets $S$ for which $G_{S \to i}$ is locally valid in the original version. However, the distinct values in $\Theta_i$ do not change. In the sample version, this modification is useful for the same reason as modification (i) of Algorithm 1. It allows us to estimate exact zeroes, without estimation error from the regression problems.

(ii) Replace $\mathrm{sib}_i(G)$ in line 3 of Algorithm 3 by $\mathrm{sib}_{i,y}(G)$. This substitution may again change the multiplicities of values in $\Theta_i$, but not the distinct values. This modification can be beneficial for two reasons. First, the algorithm may become faster, since one has to consider fewer subsets $S$ in line 3 of Algorithm 3. Second, the dimensionality of the regression problems is reduced.

(iii) Replace $\text{pa}_i(G)$ in line 5 of Algorithm 3 by $\text{pa}_{i,y}(G)$. This can be done for the same reasons as modification (ii) of Algorithm 1.

In both Algorithm 1 and Algorithm 3, one could also first determine the maximum likelihood estimator (MLE) for the estimated CPDAG, and then use this to compute the intervention effects.

We note that all modifications mentioned above are nonlocal, in the sense that they require more information about the CPDAG than the neighborhood of $X_i$, for example, about paths between $X_i$ and $Y$ [modification (i)–(iii) for Algorithm 3], or about the entire CPDAG (when using the MLE). Whether such nonlocal modifications are improvements over the local Algorithm 3 depends on the situation. When $n$ is large relative to $p$, and, hence, the CPDAG is estimated well, then the nonlocal modifications tend to behave better than Algorithm 3. On the other hand, when $n$ is not so large relative to $p$, and the CPDAG is not estimated with very high accuracy, then the nonlocal modifications tend to behave worse than Algorithm 3. The reason for the latter is that in such situations, the nonlocal modifications are more likely to use incorrect information from the estimated CPDAG than the local methods. For example, consider the situation that the true CPDAG contains one directed path from $X_i$ to $Y$, with $X_i$ and $Y$ far apart from each other. Now, if the direction of one edge on this path is reversed in the estimated CPDAG (and no other errors are made that create another directed path between $X_i$ and $Y$), then modification (i) would wrongly conclude that the effect of $X_i$ on $Y$ is zero. The local Algorithm 3 is more robust against such estimation errors, since it only requires correct estimation of the neighborhood of $X_i$ and not of the entire path between $X_i$ to $Y$.

**Acknowledgments.** We thank the referees, the Associate Editor and the Editor for their helpful and constructive comments and suggestions.

ETH ZÜRICH
SEMINAR FÜR STATISTIK
RÄMISTRASSE 101
8092 ZÜRICH
SWITZERLAND
E-MAIL: maathuis@stat.math.ethz.ch
kalisch@stat.math.ethz.ch
buhlmann@stat.math.ethz.ch